\documentclass[12pt]{article}

\usepackage{graphics}
\usepackage{url}

\begin{document}

\title{Modelling the Air Transport with\\ Complex Networks: a short review}
\author{Massimiliano Zanin$^{\textrm{a}}$ \\
Fabrizio Lillo$^{\textrm{b}}$}

\maketitle

\small
\begin{center}
  $^\textrm{a}$~\emph{Innaxis Foundation \& Research Institute, \\Jos\'e Ortega y Gasset 20, 28006, Madrid, Spain\\ 
Centre for Biomedical Technology, Polytechnic University of Madrid, Pozuelo de Alarc\'on, 28223 Madrid, Spain\\
Faculdade de Ci\^encias e Tecnologia, Departamento de Engenharia Electrot\'ecnica, Universidade Nova de Lisboa, Portugal }\\
  $^\textrm{b}$~\emph{Scuola Normale Superiore, Piazza dei Cavalieri 7, 56126, Pisa, Italy \\
Dipartimento di Fisica, viale delle Scienze I-90128, Palermo, Italy \\
Santa Fe Institute, 1399 Hyde Park Road, Santa Fe, NM 87501, USA}
\end{center}
\normalsize

\vspace{0.3cm}

\abstract{
Air transport is a key infrastructure of modern societies. In this paper we review some recent approaches to air transport, which make extensive use of theory of complex networks. We discuss possible networks that can be defined for the air transport and we focus our attention to networks of airports connected by flights. We review several papers investigating the topology of these networks and their dynamics for time scales ranging from years to intraday intervals, and consider also the resilience properties of air networks to extreme  events. Finally we discuss the results of some recent papers investigating the dynamics on air transport network, with emphasis on passengers traveling in the network and epidemic spreading mediated by air transport.
}

\section{Introduction}
\label{intro}

Transport systems, along with other infrastructures like power grids or communication networks, are fundamental
elements of our societies and economies. They guarantee the high level of mobility that we all experience, and which is vital for the cohesion of markets and for the quality of life of citizens. Moreover, transport systems enable socio-economic growth and job creation.  When such fundamental infrastructures experience a random failure or are intentionally targeted by terrorist attacks,
the whole society is severely affected. The air transport system is no exception. In $2008$ this industry generated $32$ million jobs worldwide, of which $5.5$ of them were direct, and contributed with USD $408$ billion to the global gross product \cite{Atag08}. On the other hand, its vulnerability and the consequences for citizens' mobility clearly appear when a strike or the eruption of a volcano interrupt the normal behavior of the system \cite{Bolic11,Mazzocchi10}.

The theory and application of complex networks has experienced a tremendous growth in the last decade \cite{Boccaletti06,Albert02}. In spite of its young age, the great variety of tools developed for the analysis of different topologies \cite{Costa07} has favored a better understanding of the structure and dynamics of many real-world systems \cite{Costa11}. 
Remarkably, the complex networks approach has explained the appearance of emergent phenomena in many systems composed by a large set of interacting elements.
Well known examples include social systems, e.g. the study of networks of acquaintances or the diffusion of contagious diseases \cite{Liljeros01}, the Internet \cite{Satorras01}, and applications to neural dynamics \cite{Bullmore09,Sporns04}.
It is not surprising that the complex network methodology has been successfully applied to different transportation
modes, including streets \cite{Crucitti06,Porta06}, railways \cite{Sem02}, or subways \cite{Latora02,Angeloudis06}.

In this paper we present a review of the literature related to the application of complex network theory to the air transport system. As it will be clear, several problems have been investigated so far. For instance, the description of the topological and metric structure of the network is of great importance for understanding the business strategies adopted by different airlines, for assessing passengers' mobility in the presence of direct and indirect connections, or for investigating the time evolution of air transport, while it adapts to changes in the passengers' demand and reacts to economical external forces, such as deregulation. 
Another aspect of interest is the dynamics taking place on the network. A paradigmatic example, reviewed in this paper, is the spreading of infectious diseases worldwide and the role that air transport has in enhancing the speed of epidemic propagation. 

The future presents many challenges for the air transport system and complex network theory is likely to play a more and more significant role in tackling these challenges. First of all, air transport is increasing worldwide at a very fast pace. Policy makers are aware of the fact that the current system will be at its capacity limits in few years because of the increase of traffic demand and new business challenges. For this reason, large investment programs like SESAR in Europe and SingleSky in the US have been launched.  
Also, policy makers have stressed the importance of fostering the resilience of the system, and of its capacity of recovering the required mobility after an external shock \cite{Eur11}. 
Moreover the future will require an increasing degree of integration among different transportation modes. This problem finds a natural description in terms of  multi-layer representation of complex networks \cite{Kurant06}. Clearly, all these issues are not only relevant for the air transport itself, but they have important implications for the society as a whole.

This review is organized as follows. Section \ref{constr_net} describes the main components involved in the air transport
system, which are the basis for the construction of different network representations. Section \ref{topol} reports
the most important facts about the topologies of such networks, including their dynamical evolution, and the models that have been developed to explain these characteristics. Section \ref{dyn_net} reviews the main dynamics that have been studied on top of this network and Section \ref{resil} discusses the role of network topology in the resilience and vulnerability of air transport system. Finally, Section \ref{concl} draws some final conclusions, and presents some open lines of research.

\section{Networks for the air transport}
\label{constr_net}

Many complex systems can naturally be represented by one or more networks. For instance, the Internet can be represented as a set of nodes (the webpages) connected by links (the hyperlinks) \cite{Adamic99,Albert01}. The same system can also be represented by considering the routers as nodes and their physical connections as links \cite{Vazquez02}. 

This multiple network representation property is shared also by the air transport system, and therefore one should first decide which network is investigated.  The air transport system is composed of a large number of different elements, interacting and working together.
The mobility of passengers is just the final result and it is clearly of high importance from a social point of view. Therefore it is not surprising that most analyses have been focused on the mobility of people, disregarding other technical details. 
When this point of view is followed, the construction of network is straightforward. Nodes represent airports
and a link between two nodes is created whenever there exists a direct flight between the two airports associated with the nodes.
From this point of view the airport network is the projection of the bipartite network, whose first set of nodes is composed of airports, the second set of flights, and a link exists between a flight and an airport if that flight departs or arrive to that airport. 
Clearly in this way additional sources of information, like scheduling, types of flights, or airlines, are disregarded.
The projected network of airports is naturally a directed graph, where, in general, two directed links can exist between two nodes $A$ and $B$, one describing the flights from $A$ to $B$ and one from $B$ to $A$. The projected network has also a natural weighting scheme, given by the number of flights that are present (in the investigated time period) between the two airports. These networks are termed flight networks. 

Additionally, not all flights are equivalent.  The number of available seats in each aircraft can be very different, ranging from the 50 passengers of a small regional jet, up to the 853 seats of an Airbus A380. As a consequence, it has been proposed the association of weights to links, proportional either to the frequency of connections or to the number of transported persons - see Section \ref{WNA}. As usual, from a weighted directed graph one can construct other graphs by neglecting information. For example, taking the difference between the weights  from $A$ to $B$ and from $B$ to $A$, one can construct a directed network where only one directed link exists between two nodes. By neglecting the weights one can obtain an unweighted network, where only topology matters, and by neglecting also directionality one can obtain simple binary graph. All these alternatives have been investigated in the literature. In any form, the flight network is probably the most investigated network in air traffic studies. Here we will review mostly the properties of these networks in the following of this paper.

However while complex networks are traditionally considered as static objects, it is clear that time is an important feature of any movement. This is especially true in the case of the air transport, because passengers may have to use several flights to get to their destination. If one considers only static representations of the network, there is no way of knowing the real dynamics of passengers, i.e., if one needs to wait $2$ or $10$ hours in a airport before taking the next connecting flight. In Section \ref{dyn_net} we will review some solutions that have been proposed to investigate the problem of indirect connectivity of passengers.

Another important aspect of flight networks is that they can be naturally decomposed into many subnetworks. For example, flight networks can be decomposed by considering separately one subnetwork for each airline. The analysis of the networks corresponding to a single airline has been performed, for instance, in Refs. \cite{Han04,Li04}. To the best of our knowledge, no paper has been published on the analysis of interdependencies between the subnetworks corresponding to different airlines, or of different alliances of airlines. We expect this topic will gather an increasing interest, especially because it can be considered a special case of a multi-layer representation of complex networks \cite{Kurant06}. This framework would also allow the study of the relations between the air and other transportation modes \cite{Qu10,Xu11}.

It is important to stress that networks different from the flight networks can be constructed and are likely to be very important for the understanding and modeling of air transport. The structure of the airspace is one of these cases. Nowadays aircraft do not travel along the straight line (geodesic) connecting the departure and destination airport. On the contrary, they must follow some fixed {\it airways}, defined as union of consecutive segments between pairs of navigation aids ({\it navaids} in short). While such constraints are actually imposed in order to improve safety (as it is easier to control ordered flows of aircraft) and capacity (the workload of controllers is reduced, and thus they can control a higher density of aircraft), bottlenecks may appear in some central zones of the airspace, or where several busy airways converge. Navaids can be used  to create a network, where nodes are navaids, and links represent airways. To the best of our knowledge, only very few studies have considered this type of network. For example, Cai and coworkers \cite{Cai12} investigated the Chinese air route networks.

Another example of networks considers {\it reactionary delays} and their effect on passengers. These are situations in which a flight cannot take off on time because of a delay in another flight. This occurs, for instance, because of the late arrival of the aircraft, or of the crew itself \cite{Pyrgiotis11}. Different networks may be created to study this phenomenon. Nodes may represent crews, with a link between them when they share the same aircraft. Alternatively nodes may also represent airports, connected whenever the same aircraft has to serve them in a sequential fashion. The identification of the central elements of these graphs may help in highlighting the  critical points for the dynamics of the system, and would thus allow the creation of better mitigation strategies. To the best of our knowledge, this topic has not yet been studied within this framework.

Finally we mention a very recent approach to construct networks of air traffic safety events \cite{Lillo08}. When two aircraft are too close, an automatic alarm, termed Short Term Conflict Alert (STCA), is activated and the air traffic controller is supposed to give instructions to the two pilots in order to avoid a collision. One important question is whether STCA are isolated events or whether aircraft initially involved in a STCA are likely to be involved in other STCAs with other aircraft in the near future and so on, creating a cascade of events. This possibility signals the fact that the controller suggests a local solution without forecasting unintended consequences of her instructions. By using a dataset of automatically recorded STCA, authors of  \cite{Lillo08} mapped this problem into a network of STCAs which in turn can be mapped in a network of aircraft, where two nodes (aircraft) are connected if they were involved together in a STCA. These networks shows topological regularities and might shed lights on the aircraft conflict resolution dynamics.

\section{Topological analysis}
\label{topol}

\subsection{Unweighted air transport networks}
\label{simple_proj}

The analysis of the structure of the flight network in air transport, especially when focused on individual airlines, began years before the formalization of the complex network theory. This type of analysis was motivated by the aim of defining the most efficient
structures of flights for a given airline \cite{Bania98,Alderighi07}, both in terms of yields (and, thus, profit) and of passengers' mobility. The proposed solutions can be grouped into two classes:

\begin{description}

\item[{\it Point to point}:] in this configuration, a different aircraft serves each pair of airports in the network, or at least those
pairs where the passengers' demand is enough to justify the connection. While it has the advantage of offering direct
connections to all passengers, it also requires a high number of aircraft for covering all the possible routes. For a completely connected network the number of connections increases with the square of the number of airports. This strategy of connections was common in the United States before the '70 deregulation \cite{Chou90}, and nowadays is still used by several low-cost airlines - see left panel of Fig. 1.

\item[{\it Hub-and-spoke}:] in this case, connections are structured like a chariot wheel (or a collection of such structures), in which all traffic moves along 
{\it spokes} connected to the {\it hub(s)} at the centre. While most passengers must take (at least) two different flights to reach
their destination, this strategy presents several benefits for the airline. In fact, a lower number of aircraft is required, flights usually
have a higher occupation rate, and the expansion of the network to a new airport only requires one new additional flight 
\cite{OKelly94,Berry96}. 
Today the hub-and-spokes configuration is used by most major airlines all around the world - see right panel of Fig. 1.

\end{description}

While earlier studies were mostly theoretical, the possibilities offered by the analysis of real systems through the complex networks methodology, and the ever-increasing computational capabilities of modern computers, have enabled a better understanding of the structure of real air transport networks. It is interesting to notice that some network characteristics have been confirmed in all studied networks. One important aspect of flight networks is the the fact that they show the scale-free feature. This   implies the presence of few hubs with a very high number of connections, confirming the predominance of a hub-and-spoke topology. An example can be seen in Fig. 2, which shows the cumulative probability distribution of degree of all European airports, considering only internal flights (i.e., flights whose origin or destination is outside Europe are disregarded). The right panel of Fig. 2 shows a zoom of the extreme tail of the distribution and it is clear how few airports have direct connections with a large number the destinations in the network,  performing a hub function.

\begin{table}
\caption{Example of different topological metrics of flight network, as reported in several research papers. The asterisk in the Links column indicates that the number refers to the number of flights, while in all the other cases the column reports the number of connections.}
\label{tab:Topologies}
\begin{tabular}{llrrlllllll}
\hline\noalign{\smallskip}
Country & Period & Nodes & Links & $\gamma$ & $\gamma_B$ & $L$ & $L_{rand}$ & $C$ & $C_{rand}$ & Refs. \\
\noalign{\smallskip}\hline\noalign{\smallskip}
World & 11/2000 & 3883 & 27051 & 1.0 & 0.9 & 4.4 & --- & 0.62 & 0.049 & \cite{Guimera05} \\
World & 11/2002 & 3880 & 18810 & 2.0 & --- & 4.37 & --- & --- & --- & \cite{Barrat04,Barrat05} \\
US & --- & 215 & $^*$116725 & 2.0 & --- & 1.403 & --- & 0.618 & 0.065 & \cite{LiPing03} \\
US & 10-12/2005 & 272 & 6566 & 2.63 & --- & 1.9 & 1.81 & 0.73 & 0.19 & \cite{Xu08} \\
Austria & --- & 134 & 9560 & 2.32 & --- & --- & --- & 0.206 & 0.01 & \cite{Han04} \\
China & --- & 128 & 1165 & 4.161 & --- & 2.067 & --- & 0.733 & --- & \cite{Li04} \\
China & 28/11/2007-29/3/2008 & 144 & 1018 & --- & --- & 2.23 & 1.88 & 0.69 & 0.098 & \cite{Wang11} \\
India & 12/1/2004& 79 & 442 & 2.2 & --- & 2.259 & 2.493 & 0.657 & 0.0731 & \cite{Bagler08} \\
India & 12/2010 & 84 & $^*$13909 & 0.71 & 0.54 & 2.17 & 2.55 & 0.645 & 0.18 & \cite{Sapre11} \\
Italy & 16/7-14/8/2005 & 42 & --- & 1.6 & 0.4 & 1.987 & 3.74 & 0.10 & 0.17 & \cite{Guida07} \\
Italy & 11/2005 & 42 & --- & 1.1 & 0.5 & 2.14 & 3.64 & 0.07 & 0.14 & \cite{Guida07} \\
Italy & 6/2005-5/2006 & 42 & 310 & 1.7 & 0.4 & 1.97 & --- & 0.1 & --- & \cite{Quartieri08} \\
Italy & --- & 33 & 105 & --- & --- & 1.92 & --- & 0.418 & --- & \cite{Zanin08} \\
Spain & --- & 35 & 123 & --- & --- & 1.84 & --- & 0.738 & --- & \cite{Zanin08} \\
\noalign{\smallskip}\hline
\end{tabular}
\end{table}

\begin{figure}
\begin{center}
\resizebox{0.50\columnwidth}{!}{ \includegraphics{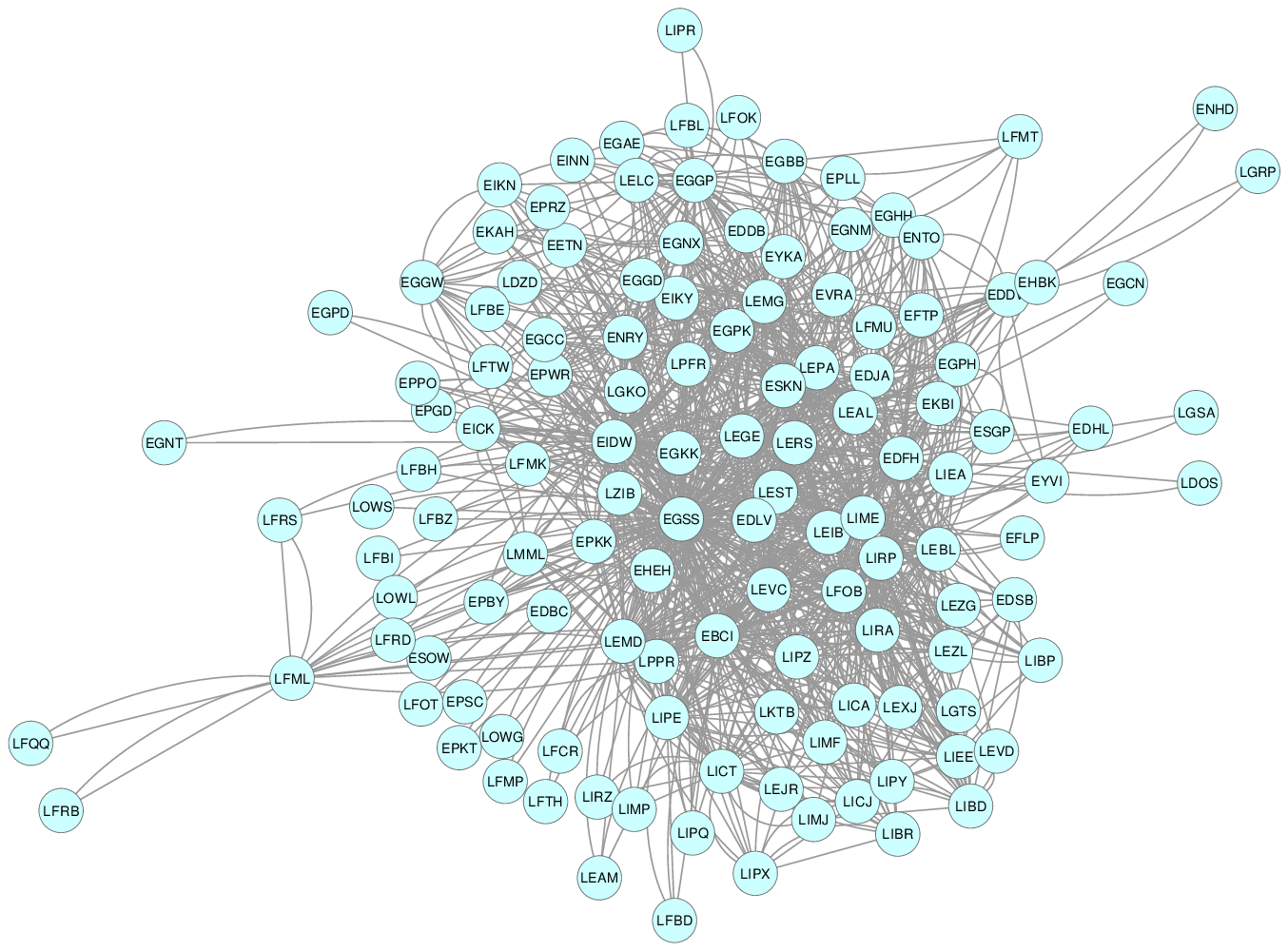} }
\resizebox{0.40\columnwidth}{!}{ \includegraphics{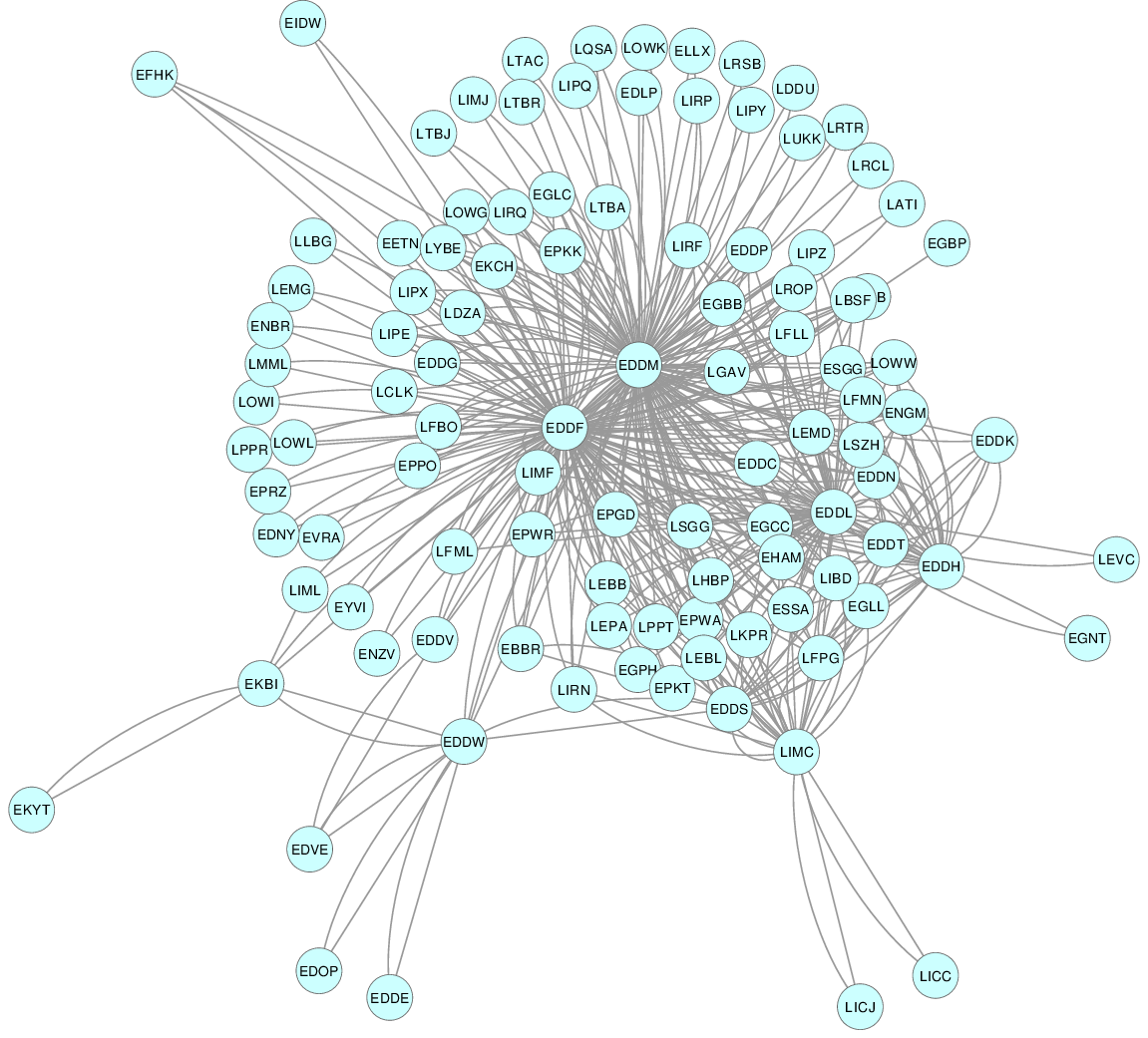} }
\caption{Representation of the networks corresponding to Ryanair (Left) and Lufthansa (Right), as at $1^{st}$ of June 2011.
Notice how the network of Lufthansa, which is a major airline, is centered around few main airports (hubs), while the structure of
Ryanair, the biggest European low-cost company, has a densely connected core.}
\end{center}
\label{fig:DLH} 
\end{figure}

\begin{figure}
\begin{center}
\resizebox{0.45\columnwidth}{!}{ \includegraphics{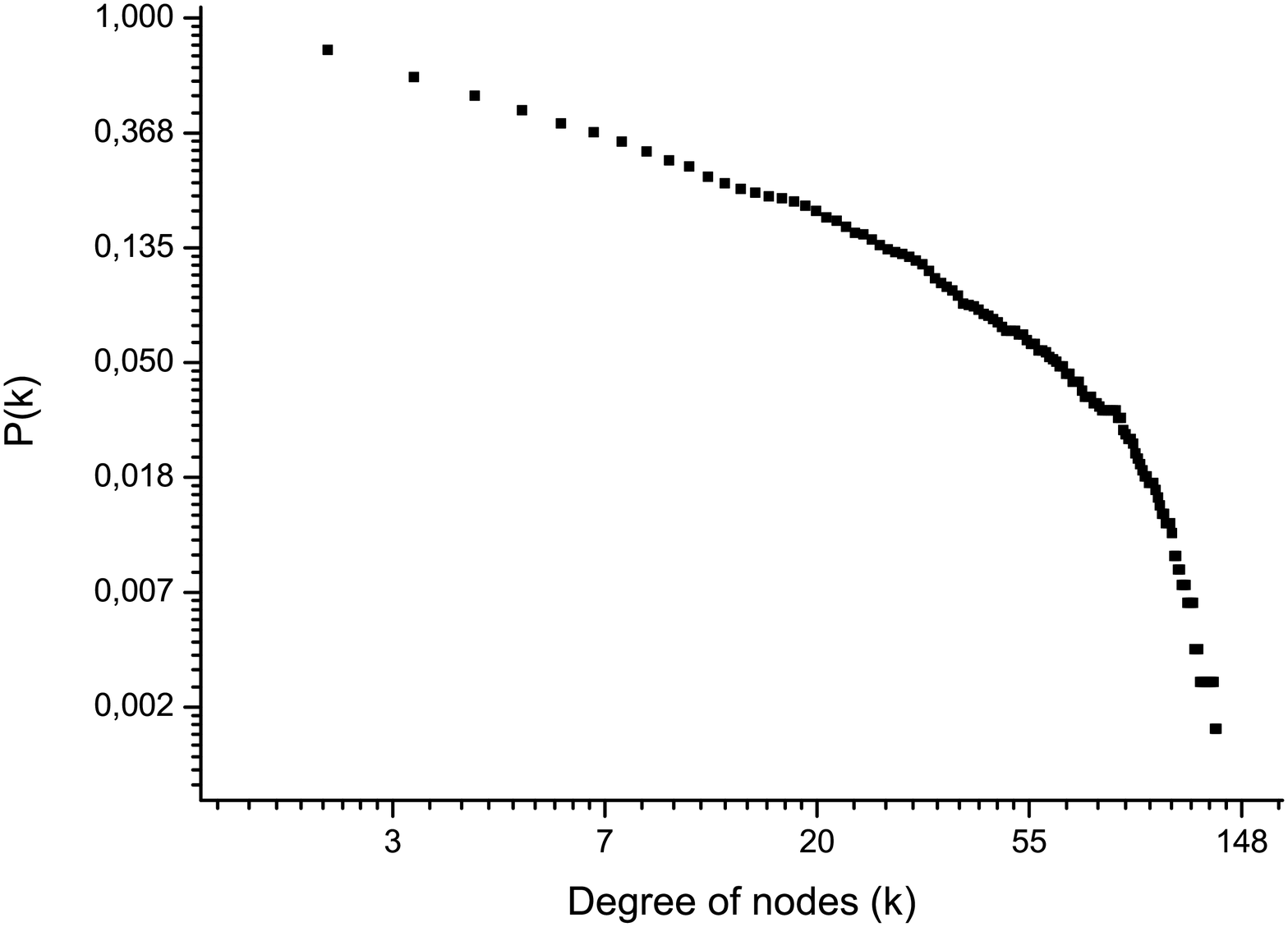} }
\resizebox{0.45\columnwidth}{!}{ \includegraphics{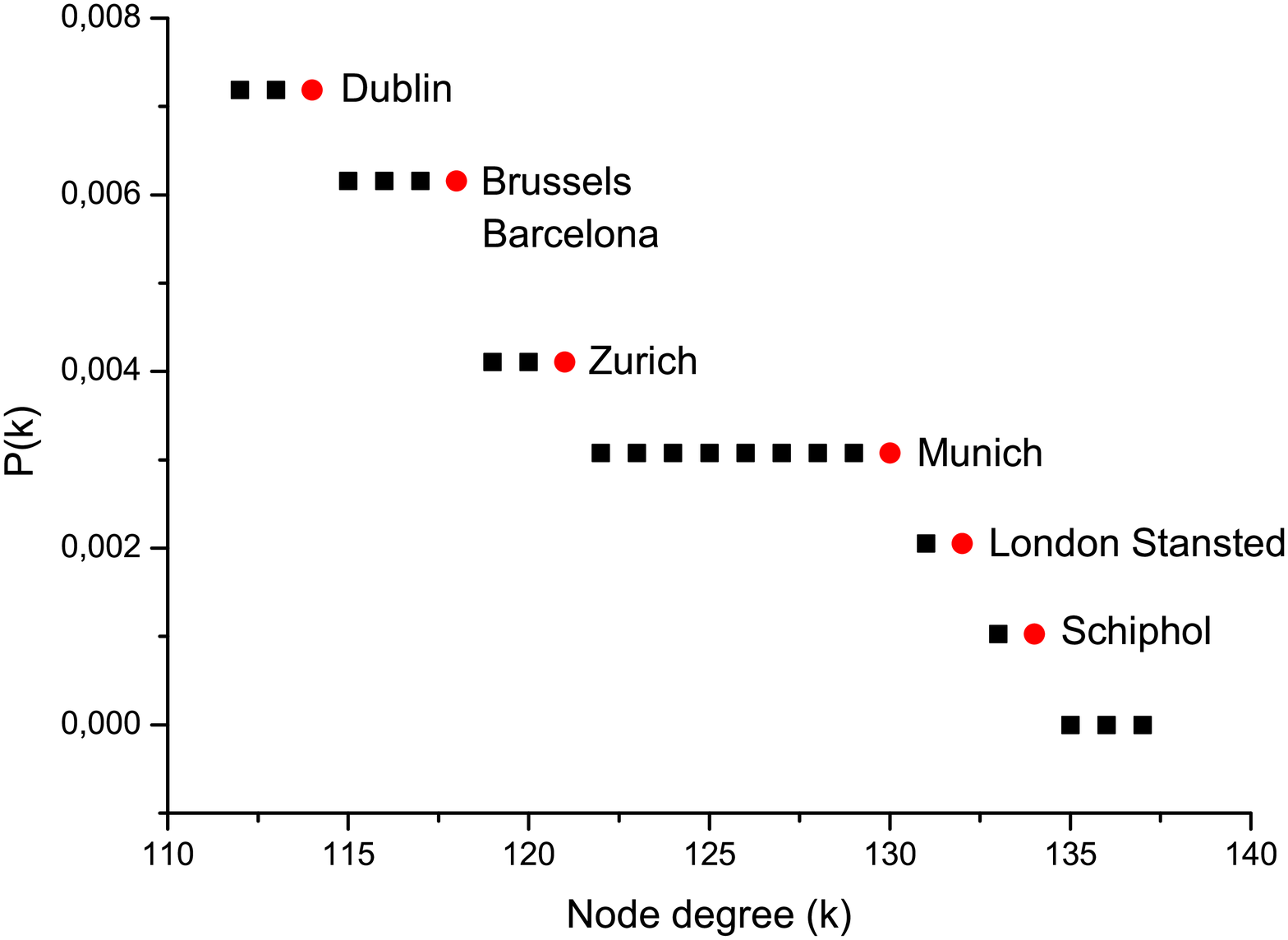} }
\caption{Cumulative probability distribution of degrees of European airports. The network has been constructed
considering only commercial (both regular and charter) flights operated between two European airports the $1^{st}$ of June 2011.
(Right) Zoom of the right part of the distribution. The hubs of the network are indicated by red circles.}
\end{center}
\label{fig:Euro} 
\end{figure}

In the literature we surveyed, different methodologies have been used for creating and analyzing flight networks. In Table \ref{tab:Topologies} we report the values of classic complex network metrics, for the air transport system of
different countries and considering an unweighted representation. The meaning of these metrics is reported below:

\begin{description}

\item[$\gamma$:] in scale-free networks \cite{Barabasi09} the asymptotic behavior of the node degree distribution has a
functional form $P(k>x) \sim x^{-\gamma}$. It has been pointed out that the real degree distribution of the worldwide flight network is a truncated power-law, i.e., it is asymptotically better explained by the function $P(k>x) \propto x^{-\gamma}f(x/k_m)$, where $f$ is an exponential truncation function and $k_m$ is a truncation parameter. The values reported in Refs. \cite{Guimera05,Barrat04} correspond to the exponent $\gamma$.

\item[$\gamma_{B}$:] the {\it betweenness} of a node is a centrality measure quantifying how important is a node for movements inside the network. Node betweenness is defined as the proportion of shortest paths,
among all possible origins and destinations, that pass through a node \cite{Freeman77}.  The exponent $\gamma_{B}$ is the exponent of a power law fit of betweenness distribution.  When the distribution of centrality is asymptotically a power-law function, a high exponent $\gamma_{B}$ indicates that few nodes are responsible for the efficient routing in the network.

\item[$L$ and $L_{rand}$:] $L$ is the mean length of shortest paths between pairs of nodes of the network, i.e.,

\begin{equation}
L = \frac{1}{{n^2 }}\sum\limits_{i,j} {d_{i,j} } 
\end{equation}

where $i$ and $j$ are two nodes of the network, $n$ the number of nodes, and $d_{ij}$ the length of the shortest (topological) 
path between nodes $i$ and $j$. The value of $L$ is usually compared with $L_{rand}$, that is the mean value obtained in
different networks that have the same number of nodes and links, but a completely random structure. These random networks  are also called
Erd\"os-R\'enyi graphs \cite{Erdos60}. It is worth noticing that the Table shows how $L$ is often lower than the corresponding $L_{rand}$, indicating, as expected, that air transport networks are engineered to efficiently reduce the number of connections needed by passengers.

\item[$C$ and $C_{rand}$:] the {\it clustering coefficient} $C$, and its randomized counterpart $C_{rand}$, measures
the number of triangles that can be found in the network \cite{Costa07}. It assesses the probability that two nodes,
which are connected to a third node, also share a direct connection.
Similarly to $L_{rand}$, $C_{rand}$ corresponds to the mean clustering coefficient of an ensemble of randomized networks.

\end{description}

The reader may easily notice how obtained values are very heterogeneous. For instance, the exponent of the degree
distribution $\gamma$ varies from $1.0$ up to $4.161$, and the clustering coefficient $C$ from $0.07$ to $0.738$. This
variability is mainly due to two factors. Firstly, there are important differences in the method used in the construction 
of the networks, which are usually not fully explained in the papers. The time window represented by the network may
be not reported \cite{Han04,Li04,Zanin08}, and no details are given about the types of flights considered (regular passengers
flights, charters, cargo flights). Secondly, most of the researches have investigated national networks, covering few tens of airports. It is well known that some  complex networks properties, such as, for instance, the scale-free distribution of degrees, are meaningful and can be correctly assessed only in large networks, where finite size effects are negligible \cite{Stumpf05}. Moreover the degree of network heterogeneity is very different if one considers a regional airport network or the worldwide network.

\subsection{Weighted network analysis}
\label{WNA}

As explained in Section \ref{simple_proj}, the analyses described above are based on unweighted projections of the 
air transport system, that is only the existence of direct connections between pairs of nodes is taken into account. On the other hand, it can be expected that the structure of frequencies of flights may unveil interesting information, especially related with the main routes
of movements chosen by passengers.

Table \ref{tab:WTopologies} shows the values of some metrics obtained for different weighted networks \cite{Newman04}. When a link between two nodes $i$ and $j$ has a weight $w_{ij}$, it is possible to calculate a weighted version of the degree of a node, called {\it strength}, as $s_i = \sum_{j} w_{ij}$. Notice that the variability observed in the metrics of the unweighted networks is amplified in the weighted network, because several variables can be used to define the value of $w_{ij}$. For example one can consider the number of flights, the number of offered seats, or the number of passengers transported, obtaining different weighted networks.

The definition of the metrics shown of Table \ref{tab:WTopologies} is here reported:

\begin{description}

\item[$\beta$:]  the relation between the strength $s$ (number of flights) of each node and its degree $k$ (number of connections) is typically well fitted by a power law $s(k) \approx k ^ {\beta}$.  This relation unveils relevant information about how capacities are distributed through the airport network. 

\item[$\beta_b$:] if one is interested in the assessment of the centrality of airports from the point of view of passengers' movements, it is possible to relate the strength of a node with its betweenness, i.e., $s(b) \approx b ^ {\beta_b}$.

\item[$\theta$:] in order to check whether there is a relation between the frequency of connections between two airports, and
their connectivity (in terms of number of destinations directly served), the weight of each link has been related with
the degrees of connected nodes, leading to $w_{ij} \approx (k_i k_j) ^ {\theta}$.

\end{description}

The main conclusion that can be drawn from Table \ref{tab:WTopologies} is that there exists a strong correlation between the
degree of a node, and the quantity of flights and passengers going through it. This fact is in agreement with the
hub-and-spoke structure of the network. The more connections a node has, the more passengers will
use that node to reach other destinations, and thus the frequencies of such connections strongly increase.

\begin{table}
\caption{Topological properties of different weighted flight networks.}
\label{tab:WTopologies}
\begin{tabular}{llllll}
\hline\noalign{\smallskip}
Country & Weigth & $\beta$ & $\beta_b$ & $\theta$ & Refs. \\
\noalign{\smallskip}\hline\noalign{\smallskip}
Worldwide & Available seats & 1.5 & 0.8 & 0.5 & \cite{Barrat04,Barrat05,Wu06} \\
US & Number of passengers & 1.8 & --- & --- & \cite{Xu08} \\
India & Number of flights & 1.43 & --- & --- & \cite{Bagler08} \\
4 European airlines & Number of flights & $(1.06 - 1.18)$ & --- & --- & \cite{Han09} \\
China & Number of flights & --- & --- & 0.5 & \cite{Li04} \\
Europe & Number of flights & 1.39 &---& ---&\cite{Lillo11}\\
\noalign{\smallskip}\hline
\end{tabular}
\end{table}

\subsection{Short and long term evolution of the network}

Although the scheduling of flights of regular airlines should be defined and published almost one year before the actual 
day of operation, the air transport system is far from being a static structure. On the contrary, airlines constantly adapt their scheduling
to changes in the passengers' demand, both on the short and long term.

On the short term, the network evolves to answer to the different needs of two groups of passengers, namely those moving for work and usually traveling from Monday to Friday, and those moving for leisure, traveling mainly during the week-end. It is well known that the first group is less sensitive to price, but highly values short travel durations. The result is that, during the week-end, the number of flights is reduced, and the network assumes a more star-like shape around the airports of cities of tourist interest.
Fig. 3 reports the evolution of two metrics, namely the mean degree $\langle k \rangle$ and the exponent
of the power-law fit of the degree distribution $\gamma$, for the Austrian and Chinese air networks at different days
of the week.

\begin{figure}
\begin{center}
\resizebox{0.50\columnwidth}{!}{ \includegraphics{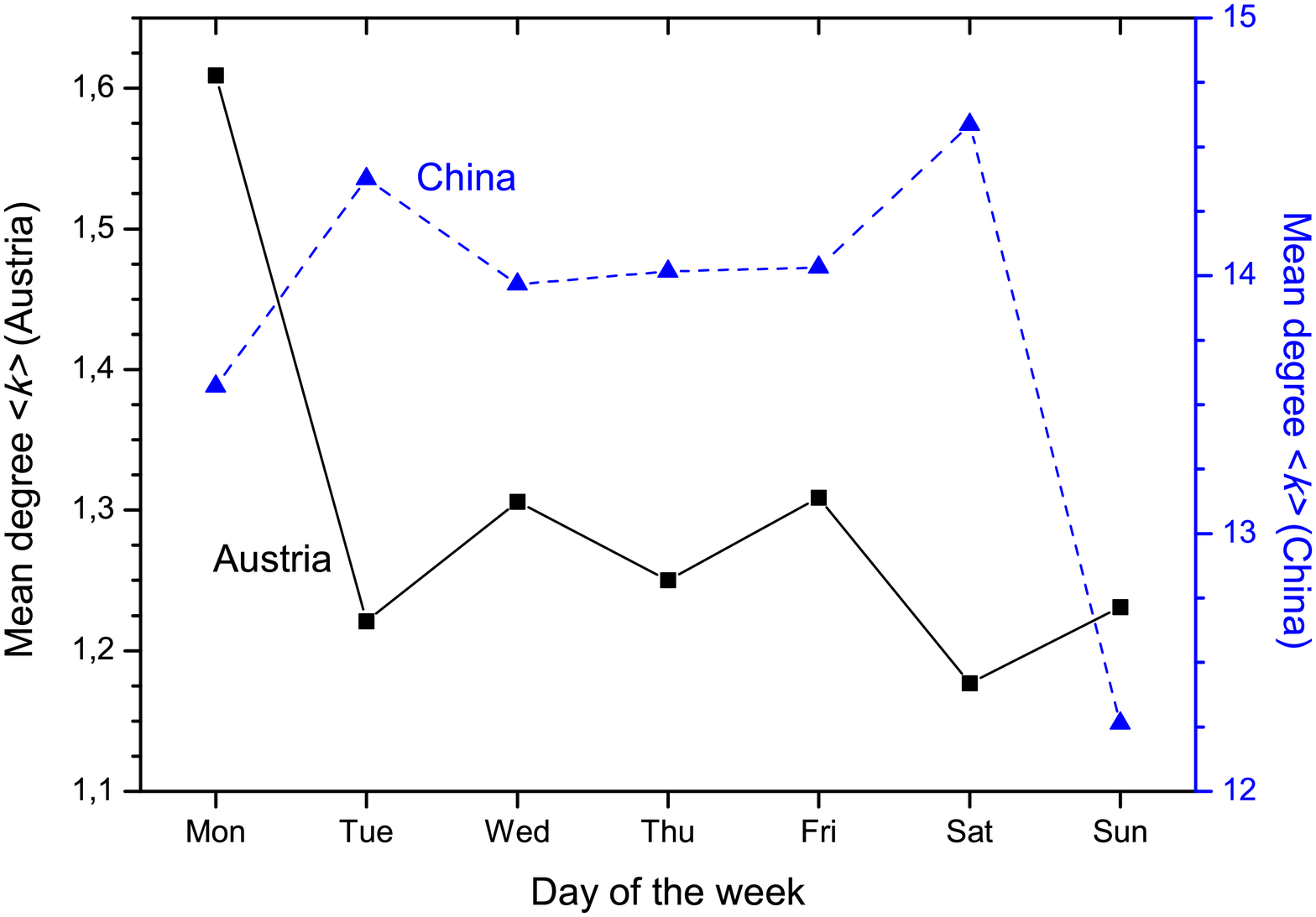} }
\resizebox{0.45\columnwidth}{!}{ \includegraphics{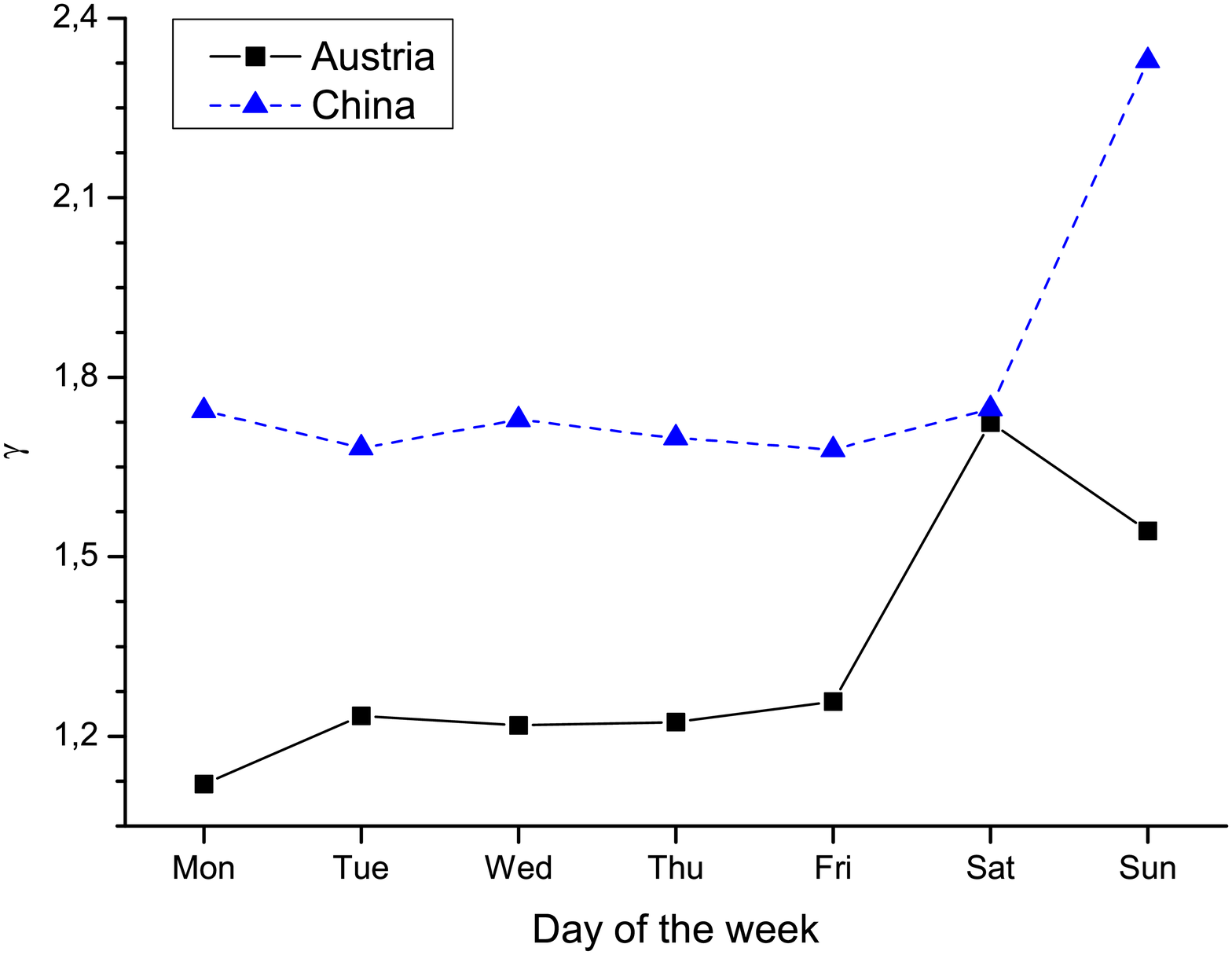} }
\caption{Evolution of the mean degree $\langle k \rangle$ (Left) and of the exponent of the degree exponential fit 
$\gamma$ (Right) for the Austrian (black squares and solid lines) and Chinese (blue triangles and dashed lines) air 
transport networks, through different days of the week. Information about the Austrian and Chinese networks have 
been obtained in Refs. \cite{Han04} and \cite{Li04} respectively.}
\end{center}
\label{fig:WeekEvol} 
\end{figure}

A long term adaptation of the network has been also studied in several papers. While short term variations are easy to predict,
long term oscillations in the demand are more complicated to forecast, because they are the result of significant changes
in macro-economical factors, of the competition between different airlines, and of the limitation imposed by policy-makers.

The evolution of the structure of the network due to changes in the regulations has been extensively studied in the case of the European deregulation started in year 1986 \cite{Barrett90} (see also Section \ref{dyn_pass}). While several papers have focused on the aeronautical and economical consequences of the deregulation
\cite{Berechman96,Betancor97,Rey03}, only a few have applied complex networks concepts to this problem.
Specifically, in Ref. \cite{Burghouwt01}, the evolution of the European air network is studied between years 1990 and 1998,
with one network created for each year.  The dynamics of some basic network metrics have been investigated. The most important of
them are the classification of airports according to their number of connections and the evolution of the number of small
and big (hubs) airports through time. Results identify two different dynamical patterns. On one side,
medium size airports have attracted most of the intra-European traffic, creating specialized internal hubs. On the other side,
the intercontinental traffic was also concentrated, but on different hubs, usually big airports. In summary, the global
structure of the network has promoted an hub-and-spoke structure, but with different hubs for internal and intercontinental
flights.

An interesting case study is represented by China. Since the economic reforms started in 1978, which slowly reduced the control of the State over the economy, the size of the air network and the number of passengers transported have increased at fast pace \cite{Jin04} (see Fig. 4). Specifically, the number of passengers grew at an average annual rate of $17\%$, mainly driven by business and tourist motivations. As a consequence of this expansion, the number of airports also increased from 69 in 1980 to 137 in 1998. As forecasted by the classical theory \cite{Chou90}, the structure of the network changed from a point-to-point to a hub-and-spoke configuration. Nevertheless this hub-and-spoke topology presents some interesting specificities \cite{Wang07}. First of all, there are three main hubs, corresponding to the headquarters of the three main national airlines, namely Beijing (Air China), Shanghai (China East) and Guangzhou (China South). The relative importance of these airports has significantly changed over time, reflecting the changes in the international scene of the corresponding cities. While Beijing was the main hub in 1990, it passed its role to Shanghai in 2005 \cite{Ma08}. Finally, these three airports form a peculiar {\it triangular subnetwork}, accounting for a $37.3\%$ of the transported passengers. These characteristics are mainly the result of regional economical and social inequalities within the Chinese territory, and of China's involvement in the world economy.
After year 2000, the main topological properties of the network have remained stationary.  However several changes have occurred,
mainly related with the appearance and extinction of small airports and routes between them \cite{Zhang10}.

\begin{figure}
\begin{center}
\resizebox{0.50\columnwidth}{!}{ \includegraphics{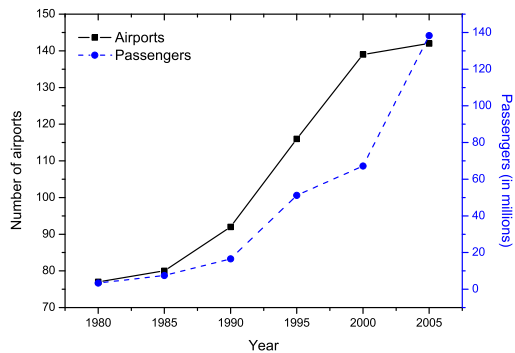} }
\caption{Evolution of the number of airports in the China's air network, and of the number of passengers from 1980 to 2005. Data correspond to Table 1 of  Ref. \cite{Wang07}.}
\end{center}
\label{fig:ChinaEvol} 
\end{figure}

To conclude this Section on the evolution of air networks, it is worth noticing Ref. \cite{Rocha09}, in which the evolution of the Brazilian air network between years 1995 and 2006 is studied. Similarly to the case of China, Brazil experienced from an important growth in its economy. Within this period, the total number of passengers transported in Brazil increased from 18 millions to 43 millions. Nevertheless, and in opposition to what found in the case of the Chinese network, the number of airports served
by major airlines strongly decreased (211 to 142), along with the mean degree (13.19 to 10.28). Therefore, while the traffic
increased, the network reduced the number of connections between secondary airports, getting closer to a pure
hub-and-spoke configuration centered around S\~{a}o Paulo, Bras\'ilia and Salvador.

\section{Dynamics on the air network}
\label{dyn_net}

Up to now we have considered the topological and metric properties of the flight network, by discussing also the dynamics {\it of the} network over short and long time scales. However the air traffic network intrinsically describes the space where something, for example passengers or goods, moves. Therefore it is natural to consider the problem of characterization, modeling, and control of the dynamics {\it on the} flight network. In this Section we review some recent approaches to dynamics on air transport network, focusing our attention to the problem of the dynamics of passengers when connection is not direct, the emergence of traffic jams, and the propagation of epidemics through the air traffic network.

\subsection{Indirect connectivity and passengers dynamics}
\label{dyn_pass}

Probably the most important entity moving on the air transport network are passengers. The structure of the air network strongly affects the capability for a passenger to reach destination B starting from destination A in the shortest possible time and in the most direct way. The transition from the point to point system to the hub and spoke system had several effects on passengers \cite{OKelly94,Berry96}. On one side, it has been argued that the hub and spoke system increased airline's efficiency and therefore, in a competitive market, it lead to lower prices. On the other side, it is clear that spoke cities risk being marginalized, being connected in a more indirect way to the rest of the system. The emergence of low cost carriers, thanks to the deregulation of the market, created cheaper opportunity of flying and created additional hub and spoke subnetworks, where hubs are different from the one used by main airlines  \cite{Dobruszkes06}. These qualitative considerations on the effect of the hub and spoke system on passengers have been investigated recently in a more quantitative way by making use of network theory, as we will show in the following. 

Many papers have focused on the dynamics of the air transport network toward an higher spatial concentration, which means that the topology of the network becomes more and more similar to a collection of star like structures. For example, several authors have investigated the change of topology of the American \cite{Goetz97} and European \cite{Burghouwt01,Button02} air transport network during the transition from the point to point to the hub and spoke structure.  However as pointed out, for example, in Ref. \cite{Burghouwt05} a hub and spoke network requires a concentration of traffic both in space and in time. Temporal concentration means that flight schedule must be organized in such a way to allow passengers to travel between two (or more) spoke cities in a relatively short time, avoiding thus long waiting time at the hubs.

In Ref. \cite{Burghouwt05} authors considered the airline network configurations in Europe between 1990 and 1999, and investigated to which extent a temporal concentration trend can be observed after deregulation, by focusing on the mechanism of wave systems. In an hub and spoke traffic network, airlines typically operate synchronized waves of flights through these hubs. The aim of such a wave-system structure is to optimize the number and quality of connections offered in the attempt to minimize waiting times for indirect connections. By comparing the flows of departing and arriving flights in an airport with an ideal type connection wave (given some transfer times), authors were able to identify the presence of waves and study how the structure and number of waves in an airport changed during the 1990s, when the air transport network became more and more similar to an hub and spoke system. Authors concluded that in the late 1990s major hubs had a clear wave system in place. For example, in Paris CDG airport six clear daily waves could be identified. The presence of these waves was then used to assess their effects on the quality of indirect connectivity. By introducing a suitable airport metric measuring the number of indirect connections weighted by transfer time and  a routing factor\footnote{The routing factor is the ratio between the actual in-flight time indirect connection and estimated in-flight time direct connection based on great circle distance. It measures the quality of the indirect connection.}, Burghouwt {\it et al.} \cite{Burghouwt05} concluded that in 1999 only  few airline hubs are highly competitive in the indirect market (essentially Frankfurt, Paris CDG, London Heathrow, and Amsterdam). Moreover they showed evidence that European airlines have increasingly adopted wave-system structures or intensified the existing structure during the 1990s. Finally, they found evidence that, given a certain number of direct flights, airports adopting a wave-system structure offer generally better indirect connections than airports without a wave-system structure.

In order to assess the role of the European hub and spoke network, Malighetti {\it et al.} \cite{Malighetti08} investigated different notion of centrality in the flight network, by considering the point of view of a passenger that must reach a destination in the shortest possible time. They first considered two purely topological measures. The first is the average shortest path length for an airport $i$, defined as the average of the minimum number of flights needed to reach an airport $j$ from airport $i$. The second metric is the betweenness of an airport $i$, i.e. the number of minimal paths that passes through airport $i$. Since there are typically many shortest paths connecting two airports (and passing through $i$), authors also defined the essential betweenness as the number of {\it unavoidable} minimal paths passing through an airport $i$. An high value of the essential betweenness indicates that the airport is a bottleneck for the traffic in the system. 
Analyzing the European air transport network in 2007, they found a large heterogeneity of average shortest path length and different values for an airport, both if one considers only European flights and if one considers also world flights. This result is somewhat expected and typically differentiates hubs for main airlines from hubs for low cost carriers.

Authors claim that, despite being interesting for the characterization of the centrality properties of the flight network, these topological measures do not help much in assessing the passengers' needs. In fact, a short path (in number of flights) can be useless for a passenger if the composing flights are unfrequent or if their scheduling makes impossible to do the connection. For this reason Malighetti {\it et al.} \cite{Malighetti08} considered a specific day and for each pair of airports they calculated the shortest travel time between them for a passenger who wants to leave at a specific time and to arrive at destination within the same day. They used scheduled flight time data and allowed for at least one hour for a flight connection. Finally they considered the minimum among the possible starting time within the day and defined the optimal path from the two airports as the one having this minimum travel time. This value is a passenger oriented metric for the distance between two airports in a given day.

It is clear that this metric takes implicitly into account the geography of Europe. In fact, most of the airports with the smallest average optimal path time are also those in the central part of Europe, in part simply because flights are shorter from there than from a peripheral airport.  Authors compared also the connectivity offered by three big alliance networks (Oneworld, Sky Team, and Star Alliance)  with that of the overall network. They found that roughly two thirds of the fastest indirect connections are not operated by the alliance system. The interpretation provided is that this could be exploited to enable a new passenger strategy of  ``self-help hubbing". However, as authors warned, this result does not take into account properly travelers' utility, since the analysis is focused on time and neglects other important variables, such as prices, number of flights, loyalty programs, etc.

A similar, but mathematically more sophisticated, approach to the indirect connectivity problem has been used in Ref. \cite{Zanin09} where {\it scheduled networks} have been introduced. A scheduled network for the air transport system is constructed starting from the ordinary flight network and by adding additional nodes on the links connecting two airports. The number of additional (secondary) nodes on a link is proportional to the traveling time needed to travel in the route associated to the link. The full network composed by primary (airports) and secondary nodes allows to compute the real time needed to go from an airport to another one, even if they are only indirected connected. The main advantage of this approach is that one can adapt several standard network metrics, such as mean path length, giant component, clustering coefficient, and tolerance to errors and attacks, to scheduled networks in order to take into account real travel time (and not just topological distance) as a metric and, more important, to take properly into account indirect connectivity. Authors then applied their method to a small sample of flight data of 40 European airports and measured the efficiency of the network as a function of the time of the day. As expected, they found a very low efficiency during the night, meaning that passengers should wait in the airport for some connections until the next morning. Moreover, they identified three peaks of connectivity during daytime, corresponding with the moments of high network connectivity (see Figure 5).

\begin{figure}
\begin{center}
\vspace{2.0cm}
\resizebox{0.50\columnwidth}{!}{ \includegraphics{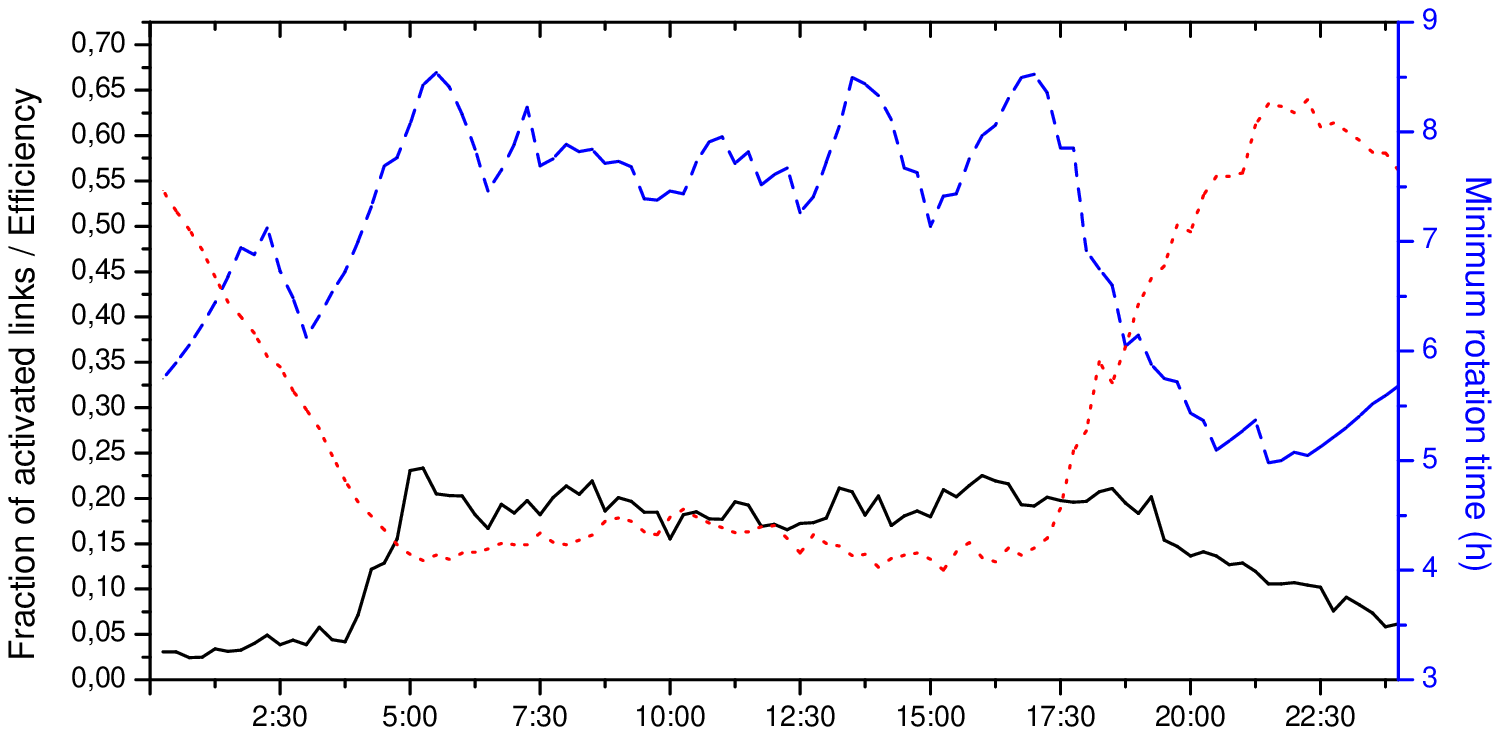} }
\caption{Evolution of three metrics for the 40 busiest European airports, at different hours of the day. The metrics are the proportion of active connections (black solid line, left scale), the efficiency of the network (blue dashed line, left scale), and the minimum {\it rotation} time, i.e., the minimum time required for a crew to return to the departure airport (red dotted line, right scale). Reprinted with permission from Ref. \cite{Zanin09}. \copyright 2009 by the American Institute of Physics.}
\end{center}
\label{fig:EvolEfficiency} 
\end{figure}

\subsection{Air traffic jams}

The commercial airline traffic is made of scheduled flights. However due to several possible reasons, such as adverse weather conditions, operational problems, and high traffic volume, the actual dynamics of flights can be drastically different from the scheduled one. Due to the strong interconnectedness of the air traffic system, it is likely that deviations from the scheduling propagate in space and time, i.e. in other airports or routes in the near future. A big engineering challenge is to design the whole system in such a way to be resilient to these shocks, i.e. to be able to return quickly to a normal state after a shock. Moreover air traffic system is a specific instance of traffic where jams can appear for no apparent reasons. A small set of recent papers started to investigate theoretically and empirically this problem.

In Ref. \cite{Lacasa09},  for example, authors proposed a network-based model of the air transport system that simulates the effect of traffic dynamics and shows the appearance of jams. Specifically, in the model a random (Erdos-Renyi) network describes the topology of a set of interconnected airports.  Each airport is characterized by an exogenously given capacity, i.e. the maximal number of aircraft per unit time that the airport can handle in an ideal situation. This ideal capacity is perturbed (diminished) by a random noise term. Moreover each link of the network is weighted and the weight measures the number of time steps that an aircraft needs in order to complete the route. A series of simple queuing rules describes the interplay between incoming and out-coming aircraft flows in an airport. Simplifying a bit, if the input flow is larger than the capacity, the output flow will be equal to the capacity and the other aircraft will remain in a queue. Only when the input flow becomes smaller than the capacity it will be possible to remove the waiting aircraft from the queue. The model is simulated with Monte Carlo methods. 

The key system indicator, $P$, is the percentage of aircraft that are not stuck in a node's queue, measured in the steady state. Note that $P$ actually measures the network's efficiency as far as it gives the flow rate which is diffusing as compared to the flow rate which is stuck. The key finding of Ref. \cite{Lacasa09} is that by increasing the aircraft density (number of aircraft), the system undergoes a phase transition. This is testified by the fact that the expected value of $P$ sharply deviates from the efficient phase $P=1$ when the aircraft density is larger than a threshold. After this threshold $P$ declines, as expected. Correspondingly, the variance of $P$, goes abruptly from zero to an high value when the aircraft density is larger than the threshold described above. These and other evidences indicate that the system undergoes a jamming transition, similarly to what observed in other traffic systems \cite{Helbing01}. One may ask whether network topology plays a role in this observation. Authors considered the topology of the (real) European air traffic system composed by $858$ airports and $11170$ flights and they  found qualitatively the same result, i.e. the emergence of a jamming phase transition for a given value of the aircraft density (see Figure 6).

\begin{figure}
\begin{center}
\resizebox{0.50\columnwidth}{!}{ \includegraphics{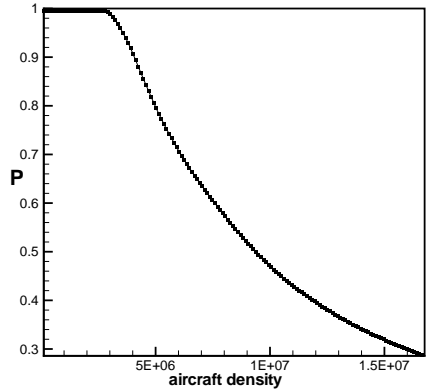} }
\caption{Phase diagram relating the percentage of aircraft  not stuck in a node's queue $P$ as a function of the network aircraft density, for the European air transport network composed of 858 nodes. Reprinted with permission from Ref. \cite{Lacasa09}. \copyright 2009 by Elsevier.}
\end{center}
\label{fig:Jamming} 
\end{figure}

\subsection{Epidemic spreading}

Another recent stream of research has considered the role of air transportation network in the propagation of global epidemics and in the assessment of its predictability. The majority of ``classical" models of epidemic spreading considers a set of individuals located very closely one to each other, so that the connections between individuals that can propagate the epidemic are short range (in space). However air transport network provides a mechanism for  long-range heterogeneous connections that can change dramatically the diffusion properties of an epidemics.

In Ref. \cite{Colizza06} Colizza {\it et al.} developed a model of a set of more than $3000$ large cities worldwide where a major airport is present. They use IATA data on the number of available seats on any given flight connection for the year 2002 and they complemented this dataset with census data on the population of the large metropolitan area served by each airport. Interestingly, they found that the number of passengers (seats) scales as the square of the population. For each city, they simulated a classical SIR epidemic model, where each individual is either susceptible (S), or infected (I), or recovered (R). Then they used the air traffic flow data to simulate the mobility of individuals from an airport to another. In this way they were able to simulate the epidemic spreading in the world and to assess the role of the air transportation network for the global pattern of emerging diseases. To this end they compared the simulations calibrated on the real air traffic data with simulations for two benchmarks. In the first one, the air traffic network is a random Erdos-Renyi graph, while in the second one the network is a graph with the same topology of the real system but fluxes and populations are taken as uniform and equal to the average value of the corresponding real variables.  In this way authors could assess the relative importance of topological (network) and metric (population, SIR rates, etc) variables by comparing simulations of the two benchmark cases to simulations of the model fully calibrated on real data.  The striking result is that the model where the topology of the air network is faithfully reproduced but the weights are homogeneous shows simulations much more similar to those of the model fully calibrated on real data than the simulations of the random graph model.  This finding strongly indicates that (at least in the framework of the model) ``the air-transportation-network properties are responsible for the global pattern of emerging diseases" \cite{Colizza06}. The same result holds if one consider the predictability of epidemic spreading. By defining a suitable measure of the sensitivity of simulations to random perturbation, Colizza {\it et al.} \cite{Colizza06} found that the predictability (or more precisely, the reproducibility or the model's sensitivity) of the random graph model is much higher than the model fully calibrated on real data. Again, the model where only the real air transport network topology is reproduced displays patterns similar to the model fully calibrated on real data. These results point out that topology of the air transport network have an important role, not only for the mobility of people, but also for the dynamics of entities that depend on human mobility.

\section{Resilience and vulnerability}
\label{resil}

In this last Section we review some results on the resilience of the air transport network, i.e., its ability to adjust its functioning prior to, during, and following internal and external disturbances, so that it can sustain required operations under both expected and unexpected conditions \cite{Hollnagel11}. For example, few preliminary studies (for example, \cite{Lillo11}) found a positive correlation between node's topological properties and typical fraction of delayed flights.
In spite of its relevance for passengers and society in general \cite{Eur11} however, little effort has been devoted to the understanding of the relationships between the topology of the air transport networks and the vulnerability of its dynamics. 

Two significant exceptions can be found in the literature. The first one, by Chi and coworkers \cite{Chi04}, analyzes how the main topological properties of the US air transport network are changed by random failures and attacks. The former effect is analyzed by de-activating airports at random, and thus simulating random disturbances like emergency situations or adverse weather. The latter is investigated by deleting the most connected nodes (as in an intentional terrorist attack). As known in more general settings \cite{Albert00},  scale-free networks are extremely resilient to random failures. However this comes at a high price, because they are also extremely vulnerable to targeted attacks. The de-activation of the $10\%$ of the most connected airports is enough to reduce the topological efficiency \cite{Latora03} of the network of $25\%$.

The resilience of the air transport system to random failures, as obtained in Ref. \cite{Chi04}, is partly aligned with the experience that we all have as passengers. In spite of multiple failures that may appear in small airports across the network, the dynamics of the system as a whole is seldom disturbed. Yet, a complementary problem is represented by those events that push the dynamics of the system far away from its normal point of operation. For instance, black swans as large strikes or the eruption of a volcano dramatically affect the performance of the system. The eruption of the Eyjafjallaj\"okull volcano in 2010 is a clear example of such random events that have larger than expected consequences.

This problem has been tackled in Ref. \cite{Wilkinson12}. Specifically, a set of random networks has been created, in which the main properties of the topology (i.e., its scale-free nature) and the spatial position of nodes have been maintained as found in the European air transport network. Results indicate that the severe disruptions observed in 2010 are explained by the geographical correlation of the disturbances (which are not, thus, completely random, as in the model of Ref. \cite{Chi04}), and by the geographical correlation of hubs, which concentrates in the centre of Europe. The proposed solution is to move some hub airports from Germany to peripheral regions. However, although this may improve the resilience against black swans, the economical consequences for airlines would probably exceed the expected benefits.

\section{Conclusions and open lines of research}
\label{concl}

In conclusions, we have presented a short review of the recent use of complex network methods for the characterization of the structure of air transport and of its dynamics. We have shown that most of the published researches have focused on the topological and metric properties of flight networks, where nodes represent airports, and links are created between pairs of them, bringing information on the presence and the frequency of flights. Specifically, these papers can be classified within three main families. Firstly, some of them propose a simple characterization of the topology of the networks, without considering their evolution through time. The recent change from a point to point to an hub-and-spoke system has triggered a series of studies, aimed at identifying and characterizing this transition, by monitoring the evolution of network's characteristics through time. Finally, some works have focused on the dynamics on the network, as, for instance, on the movement of passengers and the epidemic spreading.  However, we believe that this is just a starting point of a fruitful breeding between air transport science and complex network theory. Many different types of networks can be defined by taking into account variables or phenomena that have not been investigated with networks. Therefore, we anticipate that many interesting contributions will be published in the near future about air transport networks involving, for example, airways and navpoints, delays, safety events, crews and physical aircraft, sectors, etc.

In the near future our society is likely to face a significant growth of air traffic. Radical organizational and institutional changes are already taking place to accommodate this increase, for instance with the progressive integration of the still nationally fragmented Eu airspace management, thanks to the Single European Sky initiative and the SESAR program. The current system will drastically change, thus making a primary research area of the development of innovative network management methods and tools. We believe that complex network theory could give a significant contribution to this challenge.

\section{acknowledgement}
Authors acknowledge S. Miccich\`e, R. N. Mantegna, and S. Pozzi for useful discussion. This work has been developed as part of the activities of the ComplexWorld Network (www. complexworld.eu). Work presented therein was co-financed by EUROCONTROL on behalf of the SESAR Joint Undertaking in the context of SESAR Work Package E, project ELSA. The paper reflects only the authors' views.  EUROCONTROL is not liable for any use that may be made of the information contained therein.


\begin{thebibliography}{}

\bibitem{Atag08}
Air Transport Action Group, The economic and social benefits of air transport 2008, (2008)

\bibitem{Bolic11}
T. Bolic, Z. Sivcev, Transportation Research Record \textbf{11-3133}, (2011) 136--143

\bibitem{Mazzocchi10}
M. Mazzocchi, F. Hansstein, M. Ragona, CESifo Forum \textbf{11}, (2010) 92--100

\bibitem{Boccaletti06}
S. Boccaletti, V. Latora, Y. Moreno, M. Chavez, D.-U. Hwang, Physics Reports \textbf{424}, (2006) 175--308

\bibitem{Albert02}
R. Albert, A.-L. Barab\'asi, Rev. Mod. Phys. \textbf{74}, (2002) 47--97

\bibitem{Costa07}
L. da F. Costa, F. A. Rodrigues, G. Travieso, P. R. Villas Boas, Advances in Physics \textbf{56}, (2007) 167--242

\bibitem{Costa11}
L. da F. Costa, O. N. Oliveira, G. Travieso, F. A. Rodrigues, P. R. V. Boas, L. Antiqueira, M. P. Viana,
L. E. C. Rocha, Advances in Physics \textbf{60}, (2011) 329--412

\bibitem{Liljeros01}
F. Liljeros, R. Edling, L. A. N. Amaral, H. E. Stanley, Y. Aberg, Nature \textbf{411}, (2001) 907--908

\bibitem{Satorras01}
R. Pastor-Satorras, A. V\'azquez, A. Vespignani, Physical Review Letters \textbf{87}, (2001) 258701

\bibitem{Bullmore09}
E. Bullmore, O. Sporns, Nature Reviews Neuroscience \textbf{10}, (2009) 186--198

\bibitem{Sporns04}
O. Sporns, D. R. Chialvo, M. Kaiserc, C. C. Hilgetag, Trends in Cognitive Sciences \textbf{8}, (2004) 418--425

\bibitem{Crucitti06}
P. Crucitti, V. Latora, S. Porta, Physical Review E \textbf{73}, (2006) 036125

\bibitem{Porta06}
S. Porta, P. Crucitti, V. Latora, Physica A \textbf{369}, (2006) 853--866

\bibitem{Sem02}
P. Sen, S. Dasgupta, A. Chatterjee, P. A. Sreeram, G. Mukherjee, S. S. Manna, arXiv:cond-mat/0208535v2

\bibitem{Latora02}
V. Latora, M. Marchiori, Physica A \textbf{314}, (2002) 109-113

\bibitem{Angeloudis06}
P. Angeloudis, D. Fisk, Physica A \textbf{367}, (2006) 553--558

\bibitem{Eur11}
European Commission, Roadmap to a Single European Transport Area - Towards a 
competitive and resource efficient transport system, (2011)

\bibitem{Kurant06}
M. Kurant, P. Thiran, Physical Review Letter \textbf{96}, (2006) 138701



\bibitem{Adamic99}
L. Adamic, Lecture Notes in Computer Science \textbf{1696}, (1999) 443--452

\bibitem{Albert01}
R. Albert, H. Jeong, A.-L. Barab\'asi, Nature \textbf{401}, (2001) 130--131

\bibitem{Vazquez02}
A. Vazquez, R. Pastor-Satorras, A. Vespignani, arXiv:cond-mat/0206084v1

\bibitem{Han04}
D. D. Han, J. H. Qian, J. G. Liu, arXiv:physics/0703193v2

\bibitem{Li04}
W. Li, X. Cai, Physical Review E \textbf{69}, (2004) 046106


\bibitem{Qu10}
Y.-Q. Qu, X.-L. Xu, S. Guan, K.-J. Li, S.-J. Pan, C.-G. Gu, Y.-M. Jiang, D.-R. He, arXiv:1010.2572 (2010)

\bibitem{Xu11}
X.-L. Xu, Y.-Q. Qu, S. Guan, Y.-M. Jiang, D.-R. He, Europhysics Letters \textbf{93}, (2011) 68002

\bibitem{Cai12}
K.-Q. Cai, J. Zhang, W.-B. Du, X.-B. Cao, Chin. Phys. B \textbf{21}, (2012) 028903



\bibitem{Pyrgiotis11}
N. Pyrgiotis, K. M. Malone, A. Odoni, Transportation Research Part C, in press

\bibitem{Bania98}
N. Bania, P. W. Bauer, T. J. Zlatoper, Transportation Research Part E \textbf{34}, (1998) 53--74

\bibitem{Lillo08} F. Lillo, S. Pozzi, A. Tedeschi G. Ferrara, G. Matrella, F. Lieutaud, B. Lucat, A. Licu,
Coupling and Complexity of Interaction of STCA Networks.
Proceedings on the EUROCONTROL Conference  8th Innovative Research Workshop \& Exhibition, Bretigny-sur-Orge (France) December 1-3 2009.

\bibitem{Alderighi07}
M. Alderighi, A. Cento, P. Nijkamp, P. Rietveld, Transport Reviews \textbf{27}, (2007) 529--549

\bibitem{Chou90}
Y. H. Chou, Transportation Planning and Technology \textbf{14}, (1990) 243--258

\bibitem{OKelly94}
M. E. O'Kelly, H. J. Miller, Journal of Transport Geography \textbf{2}, (1994) 31--40

\bibitem{Berry96}
S. Berry, M. Carnall, P. T. Spiller, NBER Working Paper No. 5561, (1996)



\bibitem{Guimera05}
R. Guimer\`a, S. Mossa, A. Turtschi, L. A. N. Amaral, PNAS \textbf{102}, (2005) 7794--7799

\bibitem{Barrat04}
A. Barrat, M. Barth\'elemy, R. Pastor-Satorras, A. Vespignani, PNAS \textbf{101}, (2004) 3747--3752

\bibitem{Barrat05}
A. Barrat, M. Barth\'elemy, A. Vespignani, J. Stat. Mech. \textbf{2005}, (2005)

\bibitem{LiPing03}
C. Li-Ping, W. Ru, S. Hang, X. Xin-Ping, Z. Jin-Song, L. Wei, C. Xu, Chin. Phys. Lett. \textbf{20}, (2003) 1393--1396

\bibitem{Xu08}
Z. Xu, R. Harriss, GeoJournal \textbf{73}, (2008) 87--102.

\bibitem{Wang11}
J. Wang, H. Mo, F. Wang, F. Jin, Journal of Transport Geography \textbf{19}, (2011) 712–-721


\bibitem{Bagler08}
G. Bagler, Physica A \textbf{387}, (2008) 2972--2980

\bibitem{Sapre11}
M. Sapre, N. Parekh, Lecture Notes in Computer Science \textbf{6744}, (2011) 376--381

\bibitem{Guida07}
M. Guida, F. Maria, Chaos Solitons and Fractals \textbf{31}, (2007) 527--536

\bibitem{Quartieri08}
J. Quartieri, M. Guida, C. Guarnaccia, S. D’Ambrosio, D. Guadagnuolo, International Journal of Mathematical Models and Methods in Applied Sciences \textbf{2}, (2008) 312--316

\bibitem{Zanin08}
M. Zanin, J. M. Buld\'u, P. Cano, S. Boccaletti, Chaos \textbf{18}, (2008) 23103


\bibitem{Barabasi09}
A.-L. Barab\'asi, Science \textbf{24}, (2009) 412--413

\bibitem{Freeman77}
L. C. Freeman, Sociometry \textbf{40}, (1977) 35--41 

\bibitem{Erdos60}
P. Erd\"os, A. R\'enyi, Publications of the Mathematical Institute of the Hungarian Academy of Sciences \textbf{5}, (1960) 17--61

\bibitem{Stumpf05}
M. P. H. Stumpf, C. Wiuf, R. M. May, PNAS \textbf{102}, (2005) 4221--4224 

\bibitem{Wu06}
Z. Wu, L. A. Braunstein, V. Colizza, R. Cohen, S. Havlin, H. E. Stanley, Physical Review E \textbf{74}, (2006) 056104

\bibitem{Han09}
D. D. Han, J. H. Qian, J. G. Liu, Physica A \textbf{388}, (2009) 71--81


\bibitem{Lillo11}
F. Lillo, S. Miccich\`e, R.N. Mantegna, V. Beato, S. Pozzi,
ELSA Project: Toward a complex network approach to ATM delays analysis,
Proceedings on the INO 2011 Conference, Toulouse (France, 2011)

\bibitem{Newman04}
M. E. J. Newman, Physical Review E \textbf{70}, (2004) 056131



\bibitem{Barrett90}
S. D. Barrett, Transportation \textbf{16}, (1990) 311--327.

\bibitem{Berechman96}
J. Berechman, J. de Wit, Journal of Transport Economics and Politics \textbf{3}, (1996) 251--274

\bibitem{Betancor97}
O. Betancor, J. Campos, Proceedings of the 1997 Air Transport Research Group (ATRG) of the WCTR Society, (1997).

\bibitem{Rey03}
M. B. Rey, Journal of Air Transport Management \textbf{9}, (2003) 195--200

\bibitem{Burghouwt01}
G. Burghouwt, J. Hakfoort, Journal of Air Transport Management \textbf{7}, (2001) 311-318


\bibitem{Jin04}
F. Jin, F. Wang, Y. Liu, The Professional Geographer \textbf{56}, (2004) 471--487

\bibitem{Wang07}
J. Wang, F. Jin, Eurasian Geography and Economics \textbf{48}, (2007) 469–-480

\bibitem{Ma08}
X. Ma, M. F. Timberlake, GeoJournal \textbf{71}, (2008) 19--35

\bibitem{Zhang10}
J. Zhang, X.-B. Cao, W.-B. Du, K.-Q. Cai, Physica A \textbf{389}, (2010) 3922--3931.

\bibitem{Rocha09}
L. E. C. Rocha, J. Stat. Mech., (2009) P04020



\bibitem{Dobruszkes06}
F. Dobruszkes, Journal of Transport Geography \textbf{14}, (2006) 249--264



\bibitem{Goetz97}
A.R. Goetz and C. J. Sutton,  Journal of the Association of American Geographers \textbf{87} (1997) 238--263.

\bibitem{Button02}
K. Button, Journal of Air Transport Management \textbf{8} (2002) 177--188.

\bibitem{Burghouwt05}
G. Burghouwt and J. de Wit,  Journal of Air Transport Management \textbf{11} (2005) 185--198.

\bibitem{Malighetti08} 
P. Malighetti, S. Paleari, and R. Redondi, Journal of Air Transport Managment \textbf{14}, (2008) 53--65.

\bibitem{Zanin09}
M. Zanin, L. Lacasa, M. Cea, Chaos \textbf{19}, (2009) 023111

\bibitem{Lacasa09}
L. Lacasa, M. Cea, and M. Zanin, Physica A \textbf{388} (2009) 3948--3954.

\bibitem{Helbing01}
D. Helbing, Review of Modern Physics \textbf{73} (2001) 1067--1141.

\bibitem{Colizza06}
V. Colizza, A. Barrat, M. Barthelemy, and A. Vespignani,
Proceedings of the National Academy of Sciences USA \textbf{103} (2006) 2015--2020.




\bibitem{Hollnagel11}
E. Hollnagel, J. Pariès, D. D. Woods and J. Wreathall, \textit{Resilience Engineering in Practice} (Ashgate, 2011).

\bibitem{Chi04}
L. P. Chi, X. Cai, International Journal of Modern Physics B \textbf{18}, (2004) 2394--2400.

\bibitem{Albert00}
R. Albert, H. Jeong, A.-L. Barab\'asi, Nature \textbf{406}, (2000), 378--382

\bibitem{Latora03}
V. Latora, M. Marchiori, Europ. Phys. Journ. B \textbf{32}, (2003) 249--263.

\bibitem{Wilkinson12}
S. M. Wilkinson, S. Dunn, S. Ma, Nat Hazards \textbf{60}, (2012) 1027--1036.



% Author, Journal \textbf{Volume}, (year) page numbers
% Author, \textit{Book title} (Publisher, place year) page numbers
% etc
\end{thebibliography}
\end{document}